\documentclass[11pt,twoside]{article}
\usepackage{./asp2014}

\aspSuppressVolSlug
\resetcounters

\begin{document}

\title{Probing nearby Galactic structure through open star clusters}
\author{Joshi, Y. C.
\affil{Aryabhatta Research Institute of Observational Sciences, Nainital, India; \email{yogesh@aries.res.in}}
}

\begin{abstract}
Based on the most complete sample of Galactic open star clusters up to 1.8 kpc, we performed statistical analysis of the distribution of open cluster parameters in order to understand the Galactic structure. The geometrical characteristics of a large number of open clusters enable us to determine solar offset and scale height and distribution of reddening material in the Galactic neighbourhood. 
\end{abstract}

\section{Introduction}
The statistical analysis of a large number of star clusters plays a pivotal role to understand the Galactic structure because individual values of cluster parameters may be affected by the considerable uncertainty in their determination. Such studies have been carried out by many authors in the past (e.g., Joshi 2005, Bonatto \& Bica, 2011). In last few years, the total number of observed open star clusters has grown multi-fold which has paved way for a better understanding of the Galactic structure. Very recently, Kharchenko et al. (2013) have published a huge catalogue under their Milky Way Star Clusters (MWSC) project and it is believed to be homogeneous as same approach was used to determine the parameters for all the clusters. We here briefly examine the cluster properties using most recent catalogue on open clusters in order to understand the Galactic structure.
\section{Data}
We used MWSC catalogue that gives the physical parameters of 3208 star clusters in a series of papers published by Kharchenko et al. (2013), Schmeja et al. (2014), and Scholz et al. (2015), respectively, where PPMXL (R\"{o}ser et al. 2010) and 2MASS (Skrutskie et al. 2006) all sky catalogues were used to determine various cluster parameters. Our analysis found that MWSC catalogue may be complete only up to a distance of 1.8 kpc (Kharchenko et al. 2013, Joshi et al. 2016). After rejecting the dubious clusters in the catalogue within the completeness limit of 1.8 kpc, we used only 1218 open clusters in the present study.
\section{Reddening distribution in the Galaxy}
One of the characteristics of the interstellar extinction is its irregular
structure in the sky. Using the reddening values of 1218 open clusters, we map the extinction
variation in the sky. One of the striking feature of the map
is the clumpy behaviour of cluster distribution at some places like
$l \sim 320^\circ - 350^\circ$. In general, the clusters in the region of
$l \sim 330^\circ$  .. $0^\circ$ .. $150^\circ$ possess more extinction while the clusters
in the region of $l \sim 180^\circ$ to $270^\circ$ show lesser extinction.

We normalize the extinction by the respective distance of the clusters and a term,
$E_n(B-V)$, called normalized extinction has been evaluated as
$$E_n(B-V) = \frac{E(B-V)}{r}$$
where $r$ is the distance of cluster from the Sun. We determined the interstellar absorption $k$ as
$$ k = \frac{A_v}{r} = R\frac{E(B-V)}{r}$$
where $R$ (total-to-selective absorption) =  3.1. In order to analyse the absorption variation in some detail, we determine maximum value of absorption $k_0$ in 8 different zones of the sky in terms of longitude and corresponding distance $z_0$  from the galactic plane (GP) through the Gaussian fit in the absorption distribution. The variation of $k_0$ as a function of mean of longitude range in each zone is drawn in Fig.~1(a) which shows a sinusoidal kind of variation. A least square solution to the variation gives
$$k_0 = 1.65 + 0.73 \sin(l + 50)$$  
\vspace{-0.75cm}
$$~~~\pm0.04~\pm0.06~~~~~~~~~\pm5$$  
The absorption is therefore found to be maximum towards $l \sim 40^\circ\pm5^\circ$ and it is minimum in the direction of $l \sim 220^\circ\pm5^\circ$. 
\begin{figure}[h]
\begin{center}
\includegraphics[width=13.0cm,height=6.7cm]{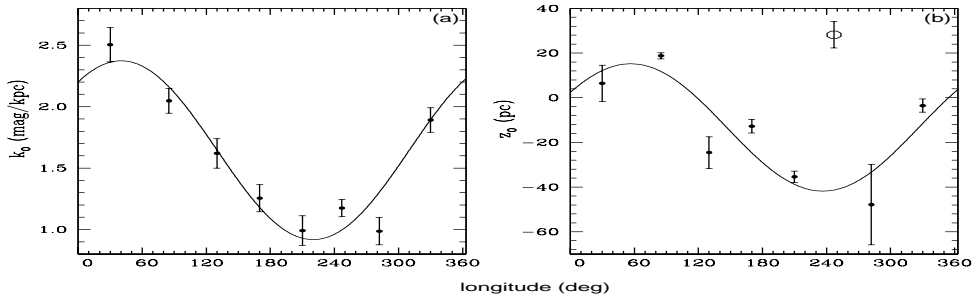}
\vspace{-2.8cm}
\caption[]{
(a). Maximum absorption $k_0$ as a function of longitude. A least square sinusoidal fit is drawn by a continuous line. (b) The height z above or below the GP at the maximum absorption $k_0$ as a function of $l$. A least square sinusoidal fit is shown by a continuous
line while dashed line shows the mean value. The lone point in the vicinity
of $l \sim 250^\circ$ drawn by open circle is omitted from the fit.}
\end{center}
\end{figure} 

\section{Solar offset}
It has long been recognized that the Sun is not precisely in the mid-plane of the Galactic disk defined by the $b = 0^\circ$ but is displaced by few parsecs to the North of GP (see, Joshi 2007) and understanding the exact value of $z_\odot$ is vital not only for the Galactic structure models but also in describing the asymmetry in the density distribution of different kind of stars in the north and south Galactic regions. Here, we determine solar offset from the number distribution of star clusters as well as distribution of reddening material around the GP.
\subsection{Number distribution}
It is normally assumed that the cluster density distribution perpendicular to
the GP could be well described in the form of a decaying exponential away from the GP, as given by
\begin{equation}
N = N_0 exp\left[-\frac{|z+z_\odot|}{z_h}\right] 
\end{equation}
where $z_\odot$ and $z_h$ are the solar offset and scale height respectively. We estimate $z_\odot$ as well as dependence of maximum value of cluster distance on $z_\odot$. This is done with different cut-off limits in $d_{max}$ using an increment of 0.3 kpc in each step and found that $z_\odot$ increases with the increasing distance. In general, it is found to be about 13 pc.
\subsection{Reddening plane}
To study the variation of maximum absorption perpendicular to the galactic
plane, we plotted the value of $z_0$ against the mean value of longitude
in Fig.~1(b).  A least square solution for a sinusoidal function gives
$$z_0 = -13 + 28 \sin(l + 33)$$  
\vspace{-0.75cm}
$$~~~~~~\pm4~~~\pm6~~~~~~~~\pm10$$  
Here, we have excluded the lone point around $230^\circ<l<265^\circ$ in drawing the
fit where $z_0$ has quite a large shift from the sinusoidal function. It shows that
the distance of the GP at maximum
absorption is symmetric around $z \sim -13$ pc which indicates that the
larger amount of reddening material is lying below the GP. A symmetric
variation around $z \sim -13$ pc means that the galactic
plane of symmetry defined by the reddening material is lying below by the same
value to the formal GP. This suggests that
solar offset is $13\pm4$ pc above the GP defined by the reddening
material. The offset of the reddening plane from the GP has
important bearing on the determination of density distribution of different
kind of stars, particularly young stars, which are closely situated near the GP.
\section{Scale height}
The observed clusters distribution perpendicular to the GP allows, in principle, the determination of the average velocity dispersion and scale height of the self gravitating matter in the disk. We fitted the number distribution of stars in the exponential profile described in Eq. (1) to estimate the scale height of the distribution of clusters which is found to be $z_h = 64\pm2$ pc. However, it is noticed that the large value of $z_h$ is influenced by the older sample of clusters which are primarily located at higher galactic latitudes resulting in a shallower wing in the $z$ distribution. If we consider only  clusters with ages $T<0.7$ Gyr neglecting the old clusters, we infer  a smaller scale height of $z_h = 60\pm2$ pc (Joshi et al. 2016).

However, a range in the cluster scale height has been reported by various authors considering cluster samples of different ages and different distance scales. We found a two-times increase in $z_h$ between the inner ($R_{GC}<=8$ kpc) and outer ($R_{GC}>8$ kpc) orbits for the older clusters, while an increase of more than five times was found between the two orbits for the youngest populations of clusters. We further note that the ratio of $z_h$ between the outer and inner orbits is almost constant for the older clusters. The increase in scale height of the disk at larger $R_{GC}$ may be explained by the decrease in gravitational potential with increasing $R_{GC}$ , or in other words, by the weaker gravitational pull exerted by the disk or disk objects as $R_{GC}$ increases. 

Considering the large sample of clusters available in the MWSC catalogue, we investigated the evolution of $z_h$ with the cluster age. We found that $z_h$ is smallest for the young clusters and largest for the old clusters. An analysis by Joshi (2007) gives the scale height of $61\pm3$ pc for the OB stars which have a similar lifetime to that of the YOCs. Since young clusters lie close to the Galactic mid-plane, like  OB stars, their scale heights are comparable. The scale height for the clusters at a mean age of about 1 Gyr reaches about 75$\pm$5 pc, which is greater than the scale height of the young clusters. Increases in $z_h$ with the age may be explained through the dynamical evolution of the Galactic disk. It is believed that the interactions of clusters with the disk and molecular clouds either evaporate the clusters or gradually disperse them from the GP over the period. Dynamical disruption is quite frequent for clusters that are closer to the center and mid-plane. The survivors reach large vertical distances from the plane. Therefore the scale height for the clusters in the Galactic disk gradually increases with the age.
\section{Summary}
Our study suggest that extinction varies sinusoidally as a function of longitude and maximum extinction is found in the direction of around $l \sim 40^\circ\pm5^\circ$. The solar offset, on the basis of reddening plane, is found to be about 13 pc above the formal GP as well as on the basis of number distribution of open clusters. We estimated an average Galactic disk scale height as $z_h = 64\pm2$ pc for the clusters  and found it strongly dependent on $R_{GC}$ and age. On an average, the scale height is more than twice larger in the outer region than in the inner region of the solar circle and, in general, increases with the age of the clusters .


\acknowledgements The author gratefully acknowledge the grant received under the joint Indo-Russian project DST/INT/RFBR/P-219 from the DST, New Delhi.



\end{document}